\documentclass[aps,prb,reprint,superscriptaddress]{revtex4-1}
\usepackage{graphicx}
\usepackage{dcolumn}
\usepackage{bm}

\begin{document}


\title{Large and homogeneous mass enhancement in the rattling-induced superconductor KOs$_2$O$_6$}


\author{Taichi Terashima}
\author{Nobuyuki Kurita}
\affiliation{National Institute for Materials Science, Tsukuba, Ibaraki 305-0003, Japan}
\author{Andhika Kiswandhi}
\author{Eun-Sang Choi}
\author{James S. Brooks}
\affiliation{National High Magnetic Field Laboratory, Florida State University, Tallahassee, FL 32310, USA}
\author{Kota Sato}
\author{Jun-ichi Yamaura}
\author{Zenji Hiroi}
\affiliation{Institute for Solid State Physics, University of Tokyo, Kashiwa, Chiba 277-8581, Japan}
\author{Hisatomo Harima}
\affiliation{Department of Physics, Graduate School of Science, Kobe University, Kobe, Hyogo 657-8501, Japan}
\author{Shinya Uji}
\affiliation{National Institute for Materials Science, Tsukuba, Ibaraki 305-0003, Japan}



\date{\today}

\begin{abstract}
We have determined the Fermi surface in KOs$_2$O$_6$ ($T_c$ = 9.6 K and $B_{c2} \sim$ 32 T) via de Haas-van Alphen (dHvA) oscillation measurements and a band structure calculation.
We find effective masses up to 26(1) $m_e$ ($m_e$ is the free electron mass), which are unusually heavy for compounds where the mass enhancement is mostly due to electron-phonon interactions.
Orbit-resolved mass enhancement parameters $\lambda_{dHvA}$ are large but fairly homogeneous, concentrated in the range 5 -- 8.
We discuss origins of the large homogeneous mass enhancement in terms of rattling motion of the K ions.
\end{abstract}


\maketitle



The alkali-metal osmium oxides $A$Os$_2$O$_6$ ($A$ = K, Rb, and Cs)\cite{Yonezawa04JPCM, Yonezawa04JPSJ_Rb, Yonezawa04JPSJ_Cs} crystallize in the cubic $\beta$-pyrochlore structure with space group $Fd\bar{3}m$ (No. 227).\cite{Yamaura06JSSC, Galati07JMC, Yamaura09SSC}
The $A$ ion is enclosed in an oversized cage formed by OsO$_6$ octahedra and vibrates in an anharmonic potential with a flat bottom, giving rise to nearly-localized low-energy anharmonic phonon modes,\cite{Kunes04PRB} i.e., rattling modes.\cite{Keppens98Nature}
The anharmonicity grows with reducing the ionic size from Cs to K, and an unusually large atomic displacement parameter $U_{iso}$ = 0.074 \AA~has been found for the K atom by an x-ray structural analysis of KOs$_2$O$_6$.\cite{Yamaura06JSSC}
Existence of low-energy rattling modes in $A$Os$_2$O$_6$ is further evidenced by various measurements.
Analyses of specific heat data indicate inclusion of Einstein modes representing rattling modes in addition to usual Debye and electronic terms is necessary to model the specific heat in $A$Os$_2$O$_6$.\cite{Bruhwiler06PRB, Hiroi07PRB, Nagao09JPSJ}
The estimated Einstein temperatures are $\theta_E$ = 22 and 61 K for $A$ = K, 66.4 K for Rb, and 75.1 K for Cs.\cite{Hiroi07PRB, Nagao09JPSJ}
The convex upward temperature dependence of resistivity\cite{Hiroi07PRB, Nagao09JPSJ} and phonon-dominated NMR relaxation rates at the K site in KOs$_2$O$_6$\cite{Yoshida07PRL} have been accounted for by considering the rattling.\cite{Dahm07PRL}
Moreover, the rattling modes have directly been observed in laser photoemission spectroscopy (PES),\cite{Shimojima07PRL} Raman scattering,\cite{Hasegawa08PRB, SchoenesPRB08, Hasegawa09JPCS} and inelastic neutron scattering.\cite{Mutka08PRB}
Finally, a mysterious first-order isomorphic phase transition observed only in KOs$_2$O$_6$ at $T_p$ = 7.6 K has been considered to be related to the rattling.\cite{Yamaura10JPSJ, Sasai10JPCM, Hattori11JPSJ}

The Sommerfeld coefficients $\gamma$ of the specific heat are estimated to be 70, 44.7, and 41.4 mJK$^{-2}$mol$^{-1}$ for $A$ = K, Rb, and Cs, respectively, yielding large mass enhancement parameters $\langle\lambda\rangle$ (= $\gamma/\gamma_{band}-1$) of 6.3, 3.38, and 2.76.\cite{Hiroi07PRB, Nagao09JPSJ}
It is generally assumed that $\langle\lambda\rangle$ due to electron-phonon interactions can not be very large.
However, since enhancement of the magnetic susceptibility in $A$Os$_2$O$_6$ is nearly absent,\cite{Hiroi07PRB, Nagao09JPSJ, Terashima10JPSJ} the observed mass enhancement is mostly due to the electron-phonon interactions, including electron-rattling ones.
As one goes from Cs to K, $\langle\lambda\rangle$ approximately doubles.
The lattice as well as electronic structure of the two compounds ($A$ = K and Cs) is basically the same.
The only significant difference is enhanced rattling motion of K in KOs$_2$O$_6$.
It can therefore be inferred that at least about half of $\langle\lambda\rangle$ in KOs$_2$O$_6$ is ascribed to the electron-rattling interactions.

$A$Os$_2$O$_6$ exhibits superconductivity below the transition temperatures of $T_c$ = 9.6, 6.3, and 3.3 K for $A$ = K, Rb, and Cs, respectively.\cite{Yonezawa04JPCM, Yonezawa04JPSJ_Rb, Yonezawa04JPSJ_Cs}
Results of specific heat,\cite{Bruhwiler06PRB, Hiroi07PRB, Nagao09JPSJ} penetration depth,\cite{Khasanov05PRB, Bonalde07PRL, Shimono07PRL} thermal conductivity,\cite{Kasahara06PRL} NMR,\cite{Magishi05PRB, Yoshida07PRL} PES,\cite{Shimojima07PRL} and scanning tunneling spectroscopy\cite{Dubois08PRL} measurements indicate fully-gapped $s$-wave superconductivity, compatible with a phonon origin of the superconductivity, possibly with moderate gap anisotropy (at most $\sim50$\%).
The specific heat jump $\Delta C$ at $T_c$ grows from $A$ = Cs to K, where $\Delta C/\gamma T_c$ of 2.87 indicates very strong coupling.\cite{Hiroi07PRB, Nagao09JPSJ}
The characteristic phonon energies $\omega_{\ln}$ contributing to the superconductivity have been estimated from the thermodynamic critical fields using a strong-coupling formula and are in good agreement with the energies of the rattling mode $\theta_E$ (for $A$ = K, the higher energy of $\theta_E$ = 61 K is used).\cite{Hiroi07PRB, Nagao09JPSJ}
This provides strong evidence that the rattling mode dominates the superconducting Cooper pairing.
The electron-phonon coupling constant relevant to the superconductivity $\lambda_{SC}$ has been estimated from a linear relation of $\lambda_{SC}$ and $T_c/\omega_{\ln}$: $\lambda_{SC}$ = 1.8, 1.33, and 0.78 for $A$ = K, Rb, and Cs, respectively.\cite{Hiroi07PRB, Nagao09JPSJ}
The large differences between $\langle\lambda\rangle$ and $\lambda_{SC}$ are intriguing.

In this Rapid Communication, we determine the Fermi surface in KOs$_2$O$_6$ via de Haas-van Alphen (dHvA) measurements and an electronic band structure calculation.
We find large effective masses up to 26(1) $m_e$ (Table I, $m_e$ is the free electron mass).
Such heavy masses are rare except for lanthanide or actinide-based heavy-fermion compounds.
Orbit-resolved mass enhancement parameters $\lambda_{dHvA}$ (= $m^*/m_{band}-1$) are in the range 5--8, consistent with the specific-heat $\langle\lambda\rangle$.
It is often argued that phonon-mediated superconductors with relatively high $T_c$ owe their high $T_c$ to strong coupling between particular phonons and particular electronic states, which will generally leads to large variation in the mass enhancement over the Fermi surface.\cite{Weber82PRB, Sadigh98PRB, Mazin03PhysicaC, Sanna07PRB}
However, by comparison with LuNi$_2$B$_2$C and MgB$_2$, we show that the mass enhancement in KOs$_2$O$_6$ can be described as fairly homogeneous over the Fermi surface.

\begin{figure}
\includegraphics{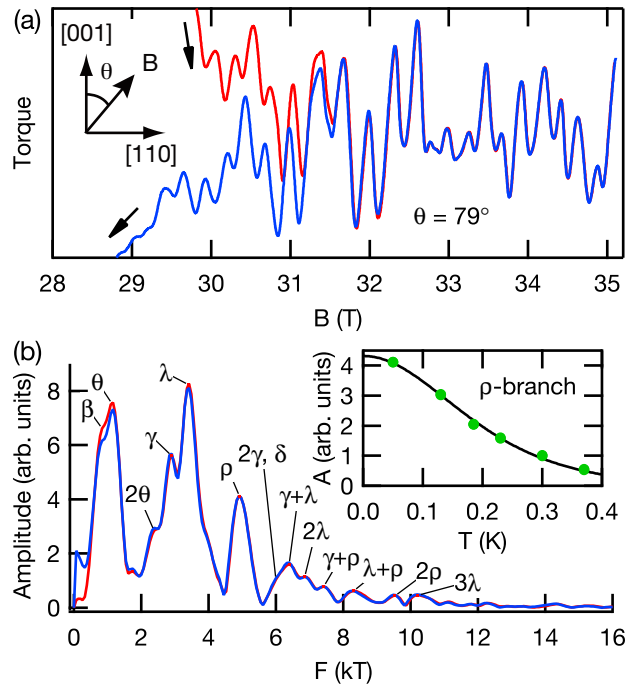}
\caption{\label{sig_ft}(color online).  (a) Magnetic torque in KOs$_2$O$_6$ at $T$ = 0.05 K for the field direction $\theta$ = 79$^\circ$.  Both field-up and field-down sweeps are shown.  Slight difference in the oscillation phase between the up and down sweeps for $B \lesssim$ 32 T is \textit{not intrinsic} but due to some experimental problem.  (b) Corresponding Fourier transforms in 1/$B$ for a field range 30.9--35.1 T, which gives a frequency resolution of $\Delta F$ = 258 T.  Fundamental dHvA frequencies labeled with Greek letters and their harmonics and combinations are resolved.  (inset) $T$-dependence of the amplitude $A$ of the frequency $\rho$.  The associated effective mass $m^*$ is estimated to be 26(1) $m_e$ from a fit to the standard Lifshitz-Kosevich formula (solid curve).}   
\end{figure}

\begin{figure}[!h]
\includegraphics[width=8.6cm]{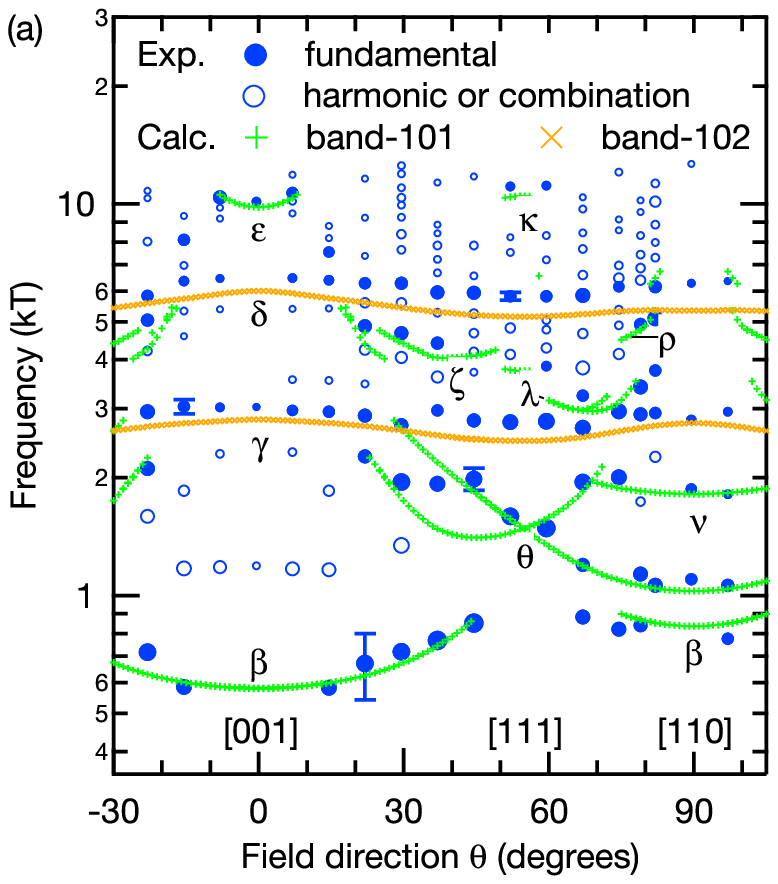}
\includegraphics[width=8.6cm]{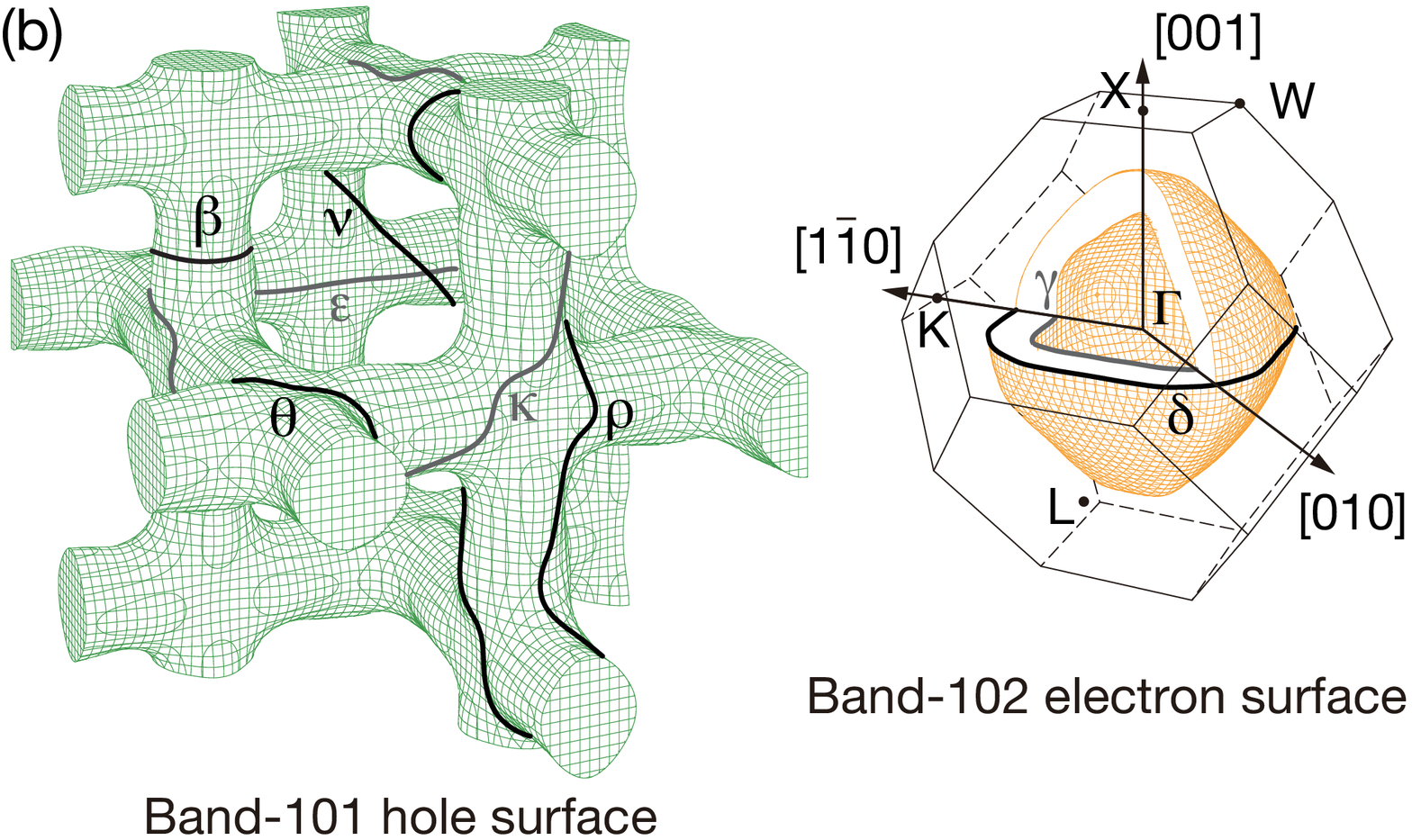}
\caption{\label{angdep}(color online).
(a) Angular dependences of the experimental and calculated dHvA frequencies.
For the experimental ones, those assigned to fundamentals are shown by solid circles, while those assigned to harmonics or combinations are shown by open circles.
The mark sizes are based on the oscillation amplitudes logarithmically.
Error bars attached to some data points are based on half of the frequency resolution, i.e., $\pm$129 T.
(b) Calculated Fermi surface of KOs$_2$O$_6$.
The hole surface (left) is shown in the repeated zone scheme.
Orthogonal pillars constituting the hole surface cross at X points.
The observed dHvA orbits are shown in the figures except for the $\zeta$ and $\lambda$ orbits, both of which are orbits on the hole surface.
$\zeta$ is a V-shaped orbit extending from an X point toward adjacent X points along the $k_x$ and $k_y$ directions. 
$\lambda$ is centered approximately at a middle of two vertically adjacent X points and extends between the two X points.
}   
\end{figure}

The KOs$_2$O$_6$ single crystal, roughly (0.2 mm)$^3$, used in this study was synthesized from mixture of KOsO$_4$ and Os as described in Ref.~\onlinecite{Hiroi07PRB}, where a residual resistivity ratio of about 300 was found for a similarly synthesized single crystal.
dHvA oscillations in magnetic torque were detected using a piezoresistive microcantilever.
A dilution refrigerator installed in a resistive magnet was used to produce temperatures $T$ down to 0.03 K in magnetic fields $B$ up to 35.1 T.
The field was rotated in the (1$\bar{1}$0) plane, and the field direction $\theta$ is measured from the [001] axis [inset of Fig.~\ref{sig_ft}(a)].
The band structure of KOs$_2$O$_6$ was calculated within the local-density approximation using a full-potential linearized augmented plane wave (FLAPW) method (TSPACE and KANSAI-06).
The obtained band structure is consistent with previous calculations\cite{Saniz04PRB, Kunes04PRB, Saniz05PRB, Hiroi07PhysC} and is very similar to that of CsOs$_2$O$_6$.\cite{Terashima08PRB}
Two bands cross the Fermi level, giving rise to the hole and electron sheets of the Fermi surface [Fig.~\ref{angdep}(b)].

\begin{table}
\caption{\label{Tab1}Experimental and calculated dHvA frequencies and effective masses.}
\begin{ruledtabular}
\begin{tabular}{ccccccc}
\multicolumn{2}{c}{} & \multicolumn{2}{c}{Experiment} & \multicolumn{2}{c}{Calculation} & \\
\cline{3-4} \cline{5-6}
\multicolumn{1}{c}{$\theta$ ($^{\circ}$)} & Branch & \multicolumn{1}{c}{$F$ (kT)} & \multicolumn{1}{c}{$m^*/m_e$} & \multicolumn{1}{c}{$F$ (kT)} & \multicolumn{1}{c}{$m_{band}/m_e$} & \multicolumn{1}{c}{$\lambda_{dHvA}$} \\
\hline
7 & $\beta$ & 0.6\footnotemark[1] & 4.7(3)\footnotemark[1] & 0.6 & 0.56 & 7.3(5)\\
7 & $\gamma$ & 3.0 & 11.5(6) & 2.8 & 1.91 & 5.0(3)\\
7 & $\delta$ & 6.5 & 14.5(9) & 5.9 & 1.89 & 6.7(5)\\
7 & $\epsilon$ & 10.7 & 20(2) & 10.4 & 3.30 & 5.0(6)\\
\\
45 & $\beta$ & 0.9 & 9.2(7) & 0.9 & 1.03 & 7.9(7)\\
45 & $\theta$ & & & 1.4 & 1.40 & \\
45 & $\theta$ & 2.0 & & 1.8 & 1.67 & \\
45 & $\gamma$ & 2.8 & 14(1) & 2.5 & 1.73 & 7.0(6)\\
45 & $\zeta$ & & & 4.1 & 3.13 & \\
45 & $\delta$ & 5.9 & 15(1) & 5.2 & 2.03 & 6.5(5)\\
\\
79 & $\beta$ & 0.8 & 8.4(9) & 0.9 & 1.01 & 7.3(9)\\
79 & $\theta$ & 1.1 & 8.8(8) & 1.1 & 0.997 & 7.8(8)\\
79 & $\nu$ & & & 1.8 & 1.73 & \\
79 & $\gamma$ & 2.9 & 14(2) & 2.7 & 1.97 & 5.8(6)\\
79 & $\lambda$ & 3.4 & 20(2) & 3.5 & 3.11 & 5.3(5)\\
79 & $\rho$ & 4.9 & 26(1) & 4.8 & 3.72 & 6.0(3)\\
79 & $\delta$ & & & 5.4 & 2.29 & \\

\end{tabular}
\end{ruledtabular}
\footnotetext[1]{Estimated from the second harmonic oscillation.}
\end{table}

Figure~\ref{sig_ft}(a) shows magnetic torque in KOs$_2$O$_6$ for $\theta$ = 79$^{\circ}$.
The field-up and field-down curves separate at a field between 31 and 32 T.
Although the field thus defined is the irreversibility field, it is comparable to $B_{c2}$ = 30.6 T\cite{Ohmichi06JPSJ} or 33 T\cite{Shibauchi06PRB} determined from penetration depth and resistivity measurements and hence we identify it with $B_{c2}$.
dHvA oscillations are clearly visible and continue below $B_{c2}$ similarly to many other type-II superconductors.\cite{Terashima97PRB, Janssen98PRB, Doiron-Leyraud07Nature}
Slight difference in the oscillation phase between the up and down sweeps, which becomes apparent for $B \lesssim B_{c2}$, is \textit{not intrinsic}.
It is not reproducible and is due to some experimental problem.\footnote{Similar phase difference between up and down sweeps were observed in some other field sweeps.  The onset fields of the difference were random and we found no correlation with $B_{c2}$.}
The Fourier transforms [Fig.~\ref{sig_ft}(b)] show several fundamental frequencies, labeled with Greek letters, and their harmonics and combinations.
The temperature dependence of the amplitude of the frequency $\rho$ is shown in the inset.
A fit to the standard Lifshitz-Kosevich formula\cite{Shoenberg84} (solid curve) indicates the associated effective mass of 26(1) $m_e$.

Figure~\ref{angdep}(a) compares the angular dependences of the experimental and calculated dHvA frequencies, and Fig.~\ref{angdep}(b) explains observed orbits including the heaviest-mass orbit $\rho$.
In our previous work,\cite{Terashima10JPSJ} when the lowest temperature was 0.6 K, we could observe only one frequency branch, $\beta$, while we have observed nearly the entire Fermi surface in this study owing to much lower measurement temperatures.
The agreement between the experimental and calculated frequencies is satisfactory.
Both $\delta$ and $\gamma$ frequencies are slightly larger than calculated as in CsOs$_2$O$_6$,\cite{Terashima08PRB} but this can not be resolved by a rigid band shift.
If the electron band is shifted up, for example, $\delta$ decreases while $\gamma$ increases.
Accordingly, we have tried no adjustment to the calculation.

We note one noticeable difference between the Fermi surfaces of KOs$_2$O$_6$ and CsOs$_2$O$_6$.
In the case of CsOs$_2$O$_6$, there are through holes connecting the inner and outer sheets of the electron surface along the $\langle111\rangle$ directions,\cite{Terashima08PRB} while in KOs$_2$O$_6$ there are no through holes and hence the inner and outer sheets are disconnected.

Table I lists the fundamental frequencies and effective masses for the three field directions where the temperature dependence was measured.
Large effective masses up to 26(1) $m_e$ and the orbit-resolved dHvA mass enhancement parameters $\lambda_{dHvA}$ of 5--8 are observed.
A comparably heavy mass of 22 $m_e$ was previously reported for a filled skutterudite compound LaFe$_4$P$_{12}$,\cite{Sugawara02PRB1} where rattling motion of La is expected.\footnote{A still heavier dHvA mass of 81 $m_e$ was reported for PrFe$_4$P$_{12}$.\cite{Sugawara02PRB1}
However, in this case, 4$f$-degrees of freedom are most likely involved in the mass enhancement and hence it can not directly be compared with the present case.}

Near absence of the Stoner enhancement and hence of the electronic mass enhancement in $A$Os$_2$O$_6$ has been evidenced by the following:
(1) The magnetic susceptibilities, which are nearly temperature independent, are almost the same for the three compounds ($A$ = K, Rb, and Cs) despite a factor of about two variation in their mass enhancements.\cite{Hiroi07PRB, Nagao09JPSJ}
(2) The Pauli paramagnetic susceptibilities, which are estimated by correcting the measured susceptibilities for the core and orbital contributions, match the band values within 30\%.\cite{Hiroi07PRB, Nagao09JPSJ}
(3) The large $B_{c2}$ of over 30 T in KOs$_2$O$_6$ is incompatible with a significant Stoner enhancement, which would reduce $B_{c2}$.\cite{Terashima10JPSJ}
The large mass enhancements are therefore for the most part due to the electron-phonon and electron-rattling interactions.

The heaviest mass found in CsOs$_2$O$_6$ is only 12 $m_e$ and $\lambda_{dHvA}$ is in the range 2.0--3.8.\cite{Terashima08PRB}
The mass enhancements are approximately doubled from Cs to K because of the enhanced rattling in KOs$_2$O$_6$, and hence roughly half of the mass enhancements in KOs$_2$O$_6$ can be ascribed to the electron-rattling interactions as already mentioned in the introduction.
It is interesting to note that electron-phonon coupling $\lambda_{ep}$ estimated from band structure calculations is about 0.8 for all the three compounds ($A$ = K, Rb, and Cs)\cite{Saniz05PRB} and hence is strikingly inconsistent with the experimental observations.

\begin{figure}
\includegraphics{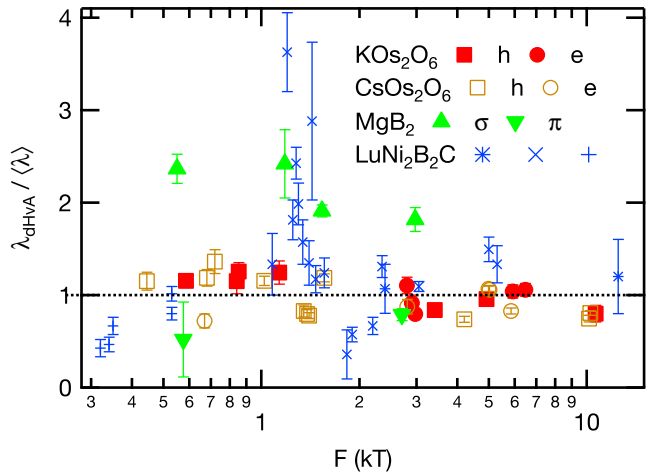}
\caption{\label{lambda}(color online).  Orbit-resolved dHvA mass enhancement parameters $\lambda_{dHvA}$ in KOs$_2$O$_6$ normalized to the Fermi-surface averaged specific-heat mass enhancement parameter $\langle\lambda\rangle$ as a function of the dHvA frequency.  Data for CsOs$_2$O$_6$,\cite{Terashima08PRB, Nagao09JPSJ} LuNi$_2$B$_2$C,\cite{Bergk08PRL, Michor95PRB} and MgB$_2$\cite{Carrington03PRL, Bouquet01PRL} are shown for comparison.  Different symbols are used for different bands as indicated by the legend in the figure.}  
\end{figure}

The determined dHvA mass enhancement parameters $\lambda_{dHvA}$ are normalized to the Fermi-surface averaged mass enhancement parameter $\langle\lambda\rangle$ from the specific heat\cite{Hiroi07PRB} and are plotted as a function of the dHvA frequency in Fig.~\ref{lambda}.
The figure also shows data for CsOs$_2$O$_6$,\cite{Terashima08PRB, Nagao09JPSJ} LuNi$_2$B$_2$C,\cite{Bergk08PRL, Michor95PRB} and MgB$_2$.\cite{Carrington03PRL, Bouquet01PRL}
Data points for different bands (Fermi surface sheets) are shown by different symbols.
In the case of MgB$_2$, the disparity between the $\sigma$ and $\pi$ bands is evident: $\lambda_{dHvA}/\langle\lambda\rangle \approx 2$ for the $\sigma$ bands (upward triangles), while $\lambda_{dHvA}/\langle\lambda\rangle < 1$ for the $\pi$ bands (downward triangles).
This is of course intimately related to the two-gap structure of the superconductivity in MgB$_2$: the superconducting energy gap $\Delta$ for the $\sigma$ bands is about three to four times larger than that for the $\pi$ bands.\cite{Bouquet01PRL}
In the case of LuNi$_2$B$_2$C, where three bands cross the Fermi level, variation in the mass enhancements is still larger.
The mass enhancement parameters not only differ from band to band but also depend on orbit orientation in one band (see crosses near $F$ = 1.1--1.4 kT distributed in the range $\lambda_{dHvA}/\langle\lambda\rangle$ = 1.3--3.6; these are data points for the same band\cite{Bergk08PRL}).
The large variation is again related to a highly anisotropic multigap structure with deep minima or possibly nodes.\cite{Terashima97PRB, Nohara97JPSJ, Boaknin01PRL}
In sharp contrast, data points of KOs$_2$O$_6$ (and also of CsOs$_2$O$_6$) are concentrated in the vicinity of $\lambda_{dHvA}/\langle\lambda\rangle=1$, and there is no clear distinction between the mass enhancements for the hole (squares) and electron (circles) bands.
This homogeneity is compatible with the limited anisotropy of the superconducting gap in $A$Os$_2$O$_6$.

It is instructive to recall here how the strongly band-dependent $\lambda$ in MgB$_2$ arises.\cite{Mazin03PhysicaC}
The $\sigma$ and $\pi$ bands in MgB$_2$ have distinct characters: the former are two-dimensional bands derived from  B $p_{xy}$ orbitals, while the latter are three-dimensional bands derived from B $p_{z}$ orbitals.
The former strongly couple with the E$_{2g}$ phonon, which is a B-B bond stretching mode, resulting in the large $\lambda$ only for the $\sigma$ bands.
In contrast, there is no clear difference in character between the hole and electron bands in KOs$_2$O$_6$: both are three dimensional, arising from Os 5$d$ and O 2$p$ orbitals.
Oxygen vibration modes will therefore couple to both bands.
Raman scattering measurements have found that those modes in $A$Os$_2$O$_6$ are more or less anharmonic and exhibit strong electron-phonon coupling.\cite{Hasegawa08PRB, Hasegawa09JPCS}
They are thus important in explaining large $\lambda$ in $A$Os$_2$O$_6$ ($\lambda$ is already unusually large in the Rb and Cs compounds without the K rattling).
For the K rattling, since electronic states of the K ion contribute little to the density of states at the Fermi level, the large electron-rattling interaction is not due to direct coupling as in MgB$_2$ between the K rattling motion and a particular electronic band, as noted in Ref.~\onlinecite{Saniz04PRB}.
It is due to the low frequencies of the rattling modes ($\lambda$ is inversely proportional to relevant phonon frequencies) and hence is basically band independent.
Thus strong band dependence of $\lambda$ is not expected in KOs$_2$O$_6$.
In addition, this indirect nature of the electron-rattling interaction is probably an important factor enabling the large mass enhancement to occur without resulting in lattice instability.

For the angular variation of $\lambda$, KOs$_2$O$_6$ is cubic and hence the angular variation is basically expected to be weak.
For the electron-rattling interaction, a more intuitive argument might be made.
The K ions sit at high symmetry points and the rattling motion is largely isotropic as evidenced by the K-ion electron density determined by x-ray diffraction.\cite{Yamaura10JPSJ}
Thus electron scattering by the rattling is basically isotropic, which, combined with the three-dimensional Fermi surface, gives nearly isotropic mass enhancement over the Fermi surface.

In summary, we have determined the Fermi surface in KOs$_2$O$_6$.
We have found the mass enhancement parameters of 5--8, which are unusually large for electron-phonon mass enhancement.
At least approximately half of the enhancement is ascribed to the electron-rattling interaction.
By comparison with MgB$_2$ and LuNi$_2$B$_2$C, we have illustrated the homogeneity of the mass enhancements in KOs$_2$O$_6$ and its relation to the relatively homogeneous superconducting gap structure.
We have discussed origins of the homogeneity in terms of the electronic band structure and rattling modes.

\begin{acknowledgments}
We thank T. Hasegawa for helpful discussion.
This work was supported by a Grant-in-Aid for Scientific Research on Innovative Areas "Heavy Electrons" (No. 23102725) of The Ministry of Education, Culture, Sports, Science, and Technology, Japan.
AK and JSB acknowledge support from NSF-DMR 1005293.
A portion of this work was performed at the NHMFL, supported by NSF Cooperative Agreement  DMR-0654118, the State of Florida, and the U.S. DoE.
\end{acknowledgments}


%

\end{document}